\documentclass{PHYEAUTH}
\usepackage{graphicx}
\usepackage{amsmath}
\usepackage{amssymb}

\begin{document}

\begin{frontmatter}

\title{Quantum phase transitions of spin chiral nanotubes}

\author[address1,crest]{Munehisa Matsumoto},
\author[address1,crest]{T\^{o}ru Sakai},
\author[address2]{Masahiro Sato},
\author[address3]{Hajime Takayama},
\author[address4]{Synge Todo}

\address[address1]{Department of Physics,
Tohoku University, Sendai 980-8578, Japan}

\address[crest]{CREST, Japan Science and Technology Agency, Kawaguchi 332-0012, Japan}

\address[address2]{Department of Physics, Tokyo Institute of Technology,
Tokyo 152-8550, Japan}

\address[address3]{Institute for Solid State Physics,
University of Tokyo,
Kashiwa 277-8581, Japan}

\address[address4]{Department of Applied Physics,
University of Tokyo, Tokyo 113-8656, Japan}

\begin{abstract}
Recently many interesting magnetic nanostructures
have been fabricated and much attention is arising
on the rich magnetic properties that originate
in the quantum effects eminent in the nanoscale world.
One of the peculiar aspects of the quantum effects
is the spin excitation gap. In the spin-$1/2$ low-dimensional
systems, the spin gap often appears when the lattice dimerization
or the frustration in the spin-spin interaction
are introduced. In the present study,
we investigate the ground-state property of
the spin-$1/2$ antiferromagnetic spin chiral nanotubes
with the spatial modulation in the spin-spin
interaction. The ground-state phase diagrams of them
are determined by observing the behavior of the
expectation value of the Lieb-Schultz-Mattis slow-twist
operator calculated by the quantum Monte Carlo method with the
continuous-time loop algorithm. We discuss
the relation between
the characteristic of the topology of the phase
diagram and the chiral vector of the nanotubes.
\end{abstract}

%
%
\end{frontmatter}

\section{Motivation}

These days the experimental techniques have made
a lot of progress and good experimentalists have
synthesized the nanoscale magnets which realize
the models that had been discussed
only theoretically before.
Among them are the BIP-TENO
as the spin-$1$ two-leg ladder~\cite{katoh},
the compound [(CuCl$_2$tachH)$_3$Cl]Cl$_2$
(tach={\it cis,trans}-1,3,5-triamino-cyclohexane)
as the triangular spin nanotube~\cite{schnack},
and the oxygen molecules adsorbed on the inner surface of the
porous material as a many-leg spin tube~\cite{shinto}.
The spin ladders~\cite{dagotto} and tubes~\cite{sato}
have attracted attention in the course of the extensive studies
on the low-dimensional magnets,
part of which started from the discussions on the Haldane gap~\cite{haldane}
of the one-dimensional spin chains. The appearance of the spin excitation gap
is one of the interesting macroscopic
quantum phenomena. In spin-$1/2$ systems,
it occurs when the frustration in the spin-spin interaction
or the lattice dimerization is introduced. A good example of the former case
is the triangular spin nanotube~\cite{sakai}
whose spontanous dimerized and gapped ground state had been
theoretically investigated~\cite{kawano} and now got
a strong possibility of an experimental realization
in the near future~\cite{schnack}.
During the course of studies on the low-dimensional systems,
the huge amount of studies on the carbon nanotubes were done
in the 90's, and the striking properties on the electric conductivity
of the chiral nanotubes were found out.
Namely, a carbon nanotube can be either
a metal or a semiconductor,
either of which is determined by the chiral vector~\cite{hamada,saito,saito2}.

With the great progress in experiments
and theories, it is now interesting in both respects to ask,
if the carbon atoms are replaced by some magnetic ions on the chiral nanotubes
and the antiferromagnetic superexchange interactions between them are introduced,
are there any relations between the chiral vector and the
magnetic properties analogous to the carbon nanotubes~\cite{yoshioka}?
The real experiments should be done in the near future.
As for the numerical experiments, we take the spin-$1/2$
Heisenberg antiferromagnets on the chiral nanotubes
made from the honeycomb strips
as the simplest systems that the quantum effects bring
about interesting properties. Remarkably, these models
always enable us
to perform numerical experiments by the quantum Monte Carlo method
without suffering from the notorious negative sign problem~\cite{troyer}.
This is not generally the case for the spin nanotubes because
they sometimes contain frustration in the spin-spin interactions~\cite{sakai,kawano}.
The details are explained in the next section.
We investigate the ground state properties
of the spin chiral nanotubes with several kinds of chiral vectors.

\section{Model}

\begin{figure}[t]
\begin{center}\leavevmode
\scalebox{0.6}{\includegraphics[width=0.75\linewidth]{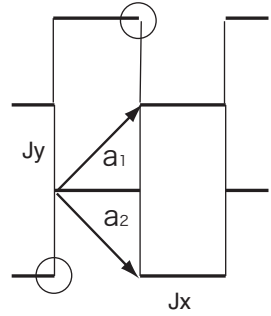}}
\caption{The lattice on which our model is defined.
The arrows ${\mathbf a}_1$ and ${\mathbf a}_2$ are the
lattice vectors. We set the coordinate system so that
the $x$-axis is horizontal.
It is topologically equivalent to the strip of the
honeycomb lattice. In this figure,
the chiral tube is made so that the
two sites encircled by a line are identical, which makes
the indices of the chiral tube $(n,m)=(3,1)$.
The definitions of $n$ and $m$ are
given in Eq.~(\ref{ham}) in the text.
}
\label{lattice}
\end{center}
\end{figure}

Our model is the spin-$1/2$ antiferromagnetic
Heisenberg tube. The tube is made from
a strip of the honeycomb lattice.
An example of the strip is shown in Fig.~\ref{lattice}.
The honeycomb strip can be seen as the coupled dimerized chains
in the strongest dimerization limit with the configuration
of the dimers in the anti-phase way
along the interchain direction.
Using the terminology
of the low-dimensional strongly correlated systems,
we call the bond along the dimerized chains
as the `leg' and that along
the interchain direction as the `rung'.
The Hamiltonian is written as follows.
\begin{eqnarray}
H&= J_x \sum_{k=1}^{n}\sum_{j=1}^{L}
\left[
\frac{1+(-1)^{j+k}}{2}
\right]
{\mathbf S}_{j,k}\cdot {\mathbf S}_{j+1,k} \nonumber \\
&+J_y\sum_{j=1}^{L}
\left(
\sum_{k=1}^{n-1}{\mathbf S}_{j,k}\cdot {\mathbf S}_{j,k+1}
+{\mathbf S}_{j,n}\cdot {\mathbf S}_{j+m,1}\right) \label{ham}
\end{eqnarray}
Here ${\mathbf S}_{j,k}$ is the spin-$1/2$ operator
with $j$ ($k$) specifying the position on the leg (rung).
The number of legs is $n$, and the number of sites
along the leg is $L$.
We impose the periodic boundary condition in both of the
rung and the leg directions such that ${\mathbf S}_{L+1,k}\equiv {\mathbf S}_{1,k}$
and ${\mathbf S}_{j,n+1}\equiv {\mathbf S}_{j+m,1}$. If $m$ is finite, chirality
is introduced in the tube structure.
The coupling constants $J_x$ and $J_y$ are positive.
We set the $x$-axis along the dimerized chains.
The ground state property in the thermodynamic limit, $L\rightarrow\infty$,
is investigated.

In our spin nanotube,
the same structure as that of the carbon nanotube
is kept when the imposed periodic boundary condition
along the rung direction satisfies the condition that
$(n+m)$ is an even number. In this case, there is no frustration
between the spin-spin interactions and we can perform
numerical experiments by the quantum Monte Carlo method
without suffering from the negative sign problem~\cite{troyer}.
The chiral vector of the tube is
defined by $p{\mathbf a_1}+q{\mathbf a_2}$, where the
lattice vectors ${\mathbf a_1}$ and ${\mathbf a_2}$ are shown in
Fig.~\ref{lattice}. In terms of our parameters, it is
written as $(p,q)=((m+n)/2,(m-n)/2)$.
%
%

\section{Method and Results}

\begin{figure}[t]
\begin{center}\leavevmode
\scalebox{0.8}{\includegraphics[width=1.0\linewidth]{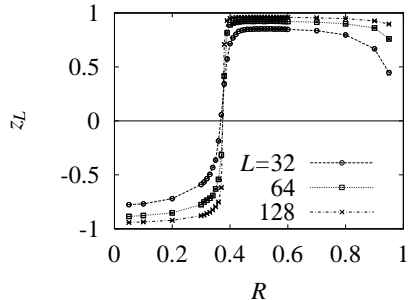}}
\caption{Behavior of the expectation value of the Lieb-Schultz-Mattis
slow-twist operator, denoted by $z_L$, for the spin nanotube with
$(n,m)=(3,1)$ plotted with respect to the parameter $R=J_y/(J_x+J_y)$.
The definitions of $(n,m)$, $J_x$, and $J_y$ are given in Eq.~(\ref{ham})
in the text.}
\label{zL_example}
\end{center}
\end{figure}

\begin{figure}[t]
\begin{center}\leavevmode
\scalebox{0.9}{\includegraphics[width=1.0\linewidth]{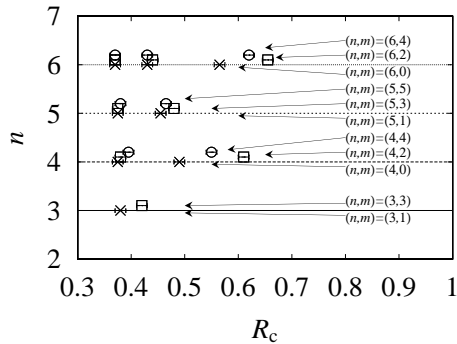}}
\caption{Distribution of the quantum critical points, $R_{\rm c}$,
of the spin nanotubes with $(n,m)=(3,1)$, $(3,3)$, $(4,0)$, $(4,2)$, $(4,4)$,
$(5,1)$, $(5,3)$, $(5,5)$, $(6,0)$, $(6,2)$, and $(6,4)$.
}
\label{qcp}
\end{center}
\end{figure}

We determine the quantum critical points of the model
and discuss their distribution in our parameter space.
The positions of them are specified by
the parameter $R\equiv J_y/(J_x+J_y)$ which connects continuously
the strong leg coupling
limit at $R=0$ to the strong rung coupling limit at $R=1$.
The phase diagram is drawn on the $R$ axis. We note that
the isotropic nanotube is realized on the point $R=0.5$.
The phase boundaries are determined by the the expectation value
of the Lieb-Schultz-Mattis slow-twist operator~\cite{nakamura,lieb}, 
which is used to detect the bond order specific to the gapped states.
For our model, it is defined as follows.
\begin{equation}
z_L=\left<
\exp\left[
i\frac{2\pi}{L}\sum_{j=1}^{L}
j\left(\sum_{k=1}^{n} S_{j,k}^{z}\right)
\right]
\right>_{0}
\end{equation}
Here $<\cdot>_{0}$ means the expectation value in the
ground state. The quantum critical points are
given by the thermodynamic limit of the zeros of $z_L$.
The observables including $z_L$ are
evaluated by the quantum Monte Carlo method
with the continuous-time loop algorithm~\cite{evertz},
which is actually a finite-temperature algorithm,
for several finite sizes. Assuming the Lorentz
invariance~\cite{chakravarty}, the temperature $T$ is
fixed to the value of $T=1/L$. The critical point
in both of the limits $L,T\rightarrow\infty$
i.e. the thermodynamic limit at zero temperature,
is determined by the crossing point of $z_L$'s
calculated for several $L$'s. An example is shown in
Fig.~\ref{zL_example} for the chiral nanotube
with $(n,m)=(3,1)$. We can identify the quantum
critical point at $R_{\rm c}=0.38\pm 0.01$.
Thus determined quantum critical points for
several chiral nanotubes are shown in Fig.~\ref{qcp}.

\section{Discussions}

The characteristic of the
resultant distribution of the
quantum critical points in Fig.~\ref{qcp}
is the difference with respect to the
parity of the number of the fully dimerized chains, $n$.
For the nanotube with even (odd) $n$, the number of
the quantum critical points is seen to be $n/2$ ($(n-1)/2$).
This reflects the well-known even-odd effect in the ground state
of uniform (i.e. non-dimerized)
spin ladders~\cite{white,dagotto} that is expected to hold
also to the spin tubes without frustration.
In this effect, the even-legged uniform systems
have an excitation gap while the odd-legged ones do not.

We sum up with a few remarks on the comparison
of the spin nanotubes with the carbon ones.
We note that $n$ is written
in terms of the chiral vector, $(p,q)$, as $n=p-q$.
The topology of the ground-state phase diagram
of the spin nanotubes is determined
by the parity of $n$, in contrast to
the carbon nanotubes where the electric property
on the point $R=0.5$ is known
by asking whether $(p-q)$ is a multiple of $3$ or not.
So far we have only determined the quantum critical points
of the spin chiral nanotubes and
discussed the topology of the ground-state
phase diagram with respect to the chiral vector.
Further investigations are now in progress.

\section*{Acknowledgements}

The authors (MM and TS)  thank
Prof. Masaki Mito and Prof. Hiroyuki Nojiri
for valuable discussions.
One of the authors (MM) would like
to express gratitude to Mr. Naoki Kobayashi,
Mr. Takashi Mesaki, and Prof. Riichiro Saito
for introducing him the physics of carbon nanotubes
and to Mr. Shinsei Ryu for useful comments.
The loop algorithm codes for the present calculations
are based on the library ``LOOPER version 2'' developed by
one of the authors (ST) and Dr. K. Kato and the codes
for parallel simulations are based on the library ``PARAPACK version 2''
also developed by one of the authors (ST).
The numerical calculations for the present work
were done on the SGI 2800 at the
Supercomputer center in the Institute for Solid
State Physics, University of Tokyo and
on the HPC Server at the Condensed Matter and
Statistical Physics Theory Group,
Tohoku University.


\begin{thebibliography}{9}
 \bibitem{katoh}
         K. Katoh, Y. Hosokoshi, K. Inoue, T. Goto:
         J. Phys. Soc. Jpn. {\bf 69} (2000) 1008.
 \bibitem{schnack}
         J. Schnack, H. Nojiri, P. K\"{o}gerler, G.~J.~T. Cooper, and L. Cronin:
         Phys. Rev. B {\bf 70} (2004) 174420.
 \bibitem{shinto}
         N. Shinto, M. Mito, T. Tajiri, H. Deguchi, S. Takagi, T. Yoshitomi, and S. Kohiki:
         the JPS autumn meeting 2004, 12aPS-115.
 \bibitem{dagotto}
         E. Dagotto and T.~M. Rice: Science {\bf 271} (1996) 618.
 \bibitem{sato}
         For the recent studies on the spin tubes, see e.g.
         M.~Sato: preprint, cond-mat/0410419 and refereces therein.
 \bibitem{haldane}
         F.~D.~M. Haldane: Phys. Lett. A {\bf 93} (1983) 464; Phys. Rev. Lett. {\bf 50} (1983) 1153.
 \bibitem{sakai}
         T. Sakai et al: presentation in this conference.
 \bibitem{kawano}
         K. Kawano and M. Takahashi: J. Phys. Soc. Jpn. {\bf 66} (1997) 4001.
 \bibitem{hamada}
         N. Hamada, S. Sawada, and A. Oshiyama: Phys. Rev. Lett. {\bf 68} (1992) 1579.
 \bibitem{saito}
         R. Saito, M. Fujita, G. Dresselhaus, and M.S. Dresselhaus:
         Phys. Rev. B {\bf 46} (1992) 1804; Appl. Phys. Lett. {\bf 60} (1992) 2204.
 \bibitem{saito2}
         R. Saito, G. Dresselhaus, and M.~S. Dresselhaus:
         {\it Physical Properties of Carbon Nanotubes},
         Imperial College Press, London (1998).
 \bibitem{yoshioka}
         The spin zigzag nanoribbons were investigated theoretically in
         H. Yoshioka: J. Phys. Soc. Jpn. {\bf 72} (2003) 2145.
 \bibitem{troyer}
         The difficulty of the negative sign problem was discussed recently
         in M. Troyer and U.-J. Wiese: preprint, cond-mat/0408370.
 \bibitem{nakamura}
	 M. Nakamura and S. Todo:
	 Phys. Rev. Lett. 89 (2002) 077204.
 \bibitem{lieb}
         E.~H. Lieb, T. Schultz, and D. Mattis: Ann. Phys. {\bf 16} (1961) 407.
 \bibitem{evertz}
	 H.~G. Evertz, G. Lana, and M. Marcu: Phys. Rev. Lett. {\bf 70} (1993) 875;
	 B.~B. Beard and U.-J. Wiese: Phys. Rev. Lett. {\bf 77} (1996) 5130;
	 S. Todo and K. Kato: Phys. Rev. Lett. {\bf 87} (2001) 047203.
 \bibitem{chakravarty}
         S. Chakravarty, B.~I. Halperin, and D.~R. Nelson:
         Phys. Rev. Lett. {\bf 60} (1988) 1057; Phys. Rev. B {\bf 39} (1989) 2344.
 \bibitem{white}
         S.~R. White, R.~M. Noack, and D.~J. Scalapino:
         Phys. Rev. Lett. {\bf 73} (1994) 886.
\end{thebibliography}
\end{document}